RUPTURE BY DAMAGE ACCUMULATION IN ROCKS

David Amitrano
LIRIGM, Université J. Fourier, Grenoble,
Maison des Geosciences, 1381 rue de la Piscine,
BP 53, 38041, Grenoble Cedex 9, France
Tel (+33) 476 82 80 85 - Fax (+33) 476 82 80 70
Email: David.amitrano@ujf-grenoble.fr

**Abstract.**
The deformation of rocks is associated with microcracks nucleation and propagation, i.e. damage. The accumulation of damage and its spatial localization lead to the creation of a macroscale discontinuity, a so-called "fault" in geological terms, and to the failure of the material, i.e., a dramatic decrease of the mechanical properties as strength and modulus. The damage process can be studied both statically by direct observation of thin sections and dynamically by recording acoustic waves emitted by crack propagation (acoustic emission). Here we first review such observations concerning geological objects over scales ranging from the laboratory sample scale (dm) to seismically active faults (km), including cliffs and rock masses (Dm, hm). These observations reveal complex patterns in both space (fractal properties of damage structures as roughness and gouge), time (clustering, particular trends when the failure approaches) and energy domains (power-law distributions of energy release bursts). We use a numerical model based on progressive damage within an elastic interaction framework which allows us to simulate these observations. This study shows that the failure in rocks can be the result of damage accumulation.

**1. Introduction.**
Deformation of rocks, when loaded at high strain rate and low temperature, involves damage processes such as microfracturing (King and Sammis, 1992; Kranz, 1983). These low scale defects induce material damage, i.e., reduced elastic and strength properties. As the crack propagation emits acoustic waves, the damage activity can be observed through seismic activity (so called acoustic emission at laboratory sample scale or micro-seismicity at rock massive scale). During the deformation process, the damage localization can lead to the nucleation of a macroscopic discontinuity (faulting) associated with a dramatic stress release which characterizes brittle behavior (Cox and Meredith, 1993; Jaeger and Cook, 1979; Lockner et al., 1991; Scholz, 1990). The change of loading conditions (reduced strain rate, increased temperature or confining pressure) induces a change of the macroscopic behavior, which becomes progressively more ductile (absence of macroscopic stress drop) and on damage repartition, which becomes more diffuse (Jaeger and Cook, 1979; Kranz, 1983; Menendez et al., 1996). The identification of the parameters controlling both the macroscopic behavior and the spatial damage repartition is an important topic for numerous geomechanics domains (underground excavation, slope instability, dams, earth-crust seismicity), because it determines the ability to predict dramatic ruptures (rockbursts, rock masses collapse, earthquakes, etc.).
The damage localization process in rocks has been often modeled to consider either a discontinuous media containing propagating cracks (Costin, 1983; Cowie et al., 1993; Li et al., 2000; Scavia, 1995) or a continuous material subject to a bifurcation phenomena (Rice, 1975; Rudnicki and Rice, 1975). An intermediary approach consists in considering the material to be continuous at mesoscale. The cracking is taken into account through elastic damage (reduction in the apparent elastic modulus). In this way, using local progressive damage and elastic interaction, previous attempts succeeded in modelling either macroscopic plasticity (Zapperi et al., 1997b) or macroscopic brittleness (Tang, 1997; Tang and Kaiser, 1998). Amitrano et al. (1999), using a local scalar damage formulation associated with a tensorial elastic interaction model, succeeded in switching continuously from macroscopic plasticity, with diffuse damage, to macroscopic brittleness, with localised damage. After these results, the brittle-ductile and localized-diffuse transitions appear to be controlled by a single parameter, the internal friction. These numerical results appear to be in good agreement with laboratory experiments (Amitrano, 2003) and earthcrust observations (Gerstenberger et al., 2001; Mori and Abercombie, 1997; Schorlemmer and Wiemer, 2005; Sue et al., 2002). In this paper we first review experimental observations of failure in rocks, distinguishing static observation of damage structure and dynamic observation of damage dynamics. Then we use a numerical model of progressive damage to show how fracture may result from damage accumulation.

## 2. Experimental observations of damage in rocks
### 2.1 Damage structure.
During the first steps of the loading of initially intact rocks, microfracturing appears to be homogeneously distributed in the whole material and mainly in mode I. As microfracturing progresses, cooperative interactions of cracks take place and lead to the coalescence of microcracks and the initiation of a macroscopic fracture which is macroscopically in mode II (Amitrano and Schmittbuhl, 2002; Costin, 1983; Kranz, 1983; Reches and Lockner, 1994; Schulson et al., 1999). Such a coalescence process has been experimentally observed by acoustic emission source location (Lockner and Byerlee, 1991). After failure, or when the discontinuity already exists, deformation is localized along the rupture band. Low-scale observations reveal that the rupture zone or shear band is made of granular material (called gouge or cataclasis in geological terms), filled in between two rupture surfaces. The different aspects of damage -- cracks, rupture surface, gouge -- that result from the deformation process, can be observed either at the field scale (natural faults) or at the laboratory sample scale (e.g. Keller et al., 1997; Wibberley et al., 2000). Shear deformation occurs both on the rupture surface and within the gouge layer involving friction surface erosion (e.g. Wang and Scholz, 1994) and grain fracturing (Biarez and Hicher, 1997; Michibayashi, 1996). The latter reduces particle size as shear progresses. Thin particles might form subshear bands as observed both at laboratory scale or at field scale (Amitrano and Schmittbuhl, 2002; Boullier et al., 2004; Lin, 1999; Mair et al., 2000; Menendez et al., 1996; Moore et al., 1989). Each aspect of the damage process during fracturing reveals scaling invariances (King and Sammis, 1992; Turcotte, 1992). Power law scaling is found for: crack lengths, crack spatial distributions (Hirata et al., 1987;

Velde et al., 1993), rupture surface roughness (Bouchaud, 1997; Bouchaud et al., 1990; Brown and Scholz, 1985; Schmittbuhl et al., 1993; Schmittbuhl et al., 1995), and grain-size distribution of the gouge (Amitrano and Schmittbuhl, 2002; Boullier et al., 2004; Marone and Scholz, 1989; Sammis and Biegel, 1989; Weiss and Gay, 1998). The self affine properties of fracture roughness is widely observed in mode I fracture obtained by traction. They have been also observed for fracture resulting from damage localization which is macroscopically mode II (Amitrano and Schmittbuhl, 2002). The power-law distribution of grain size has been also observed for gouge sampled within the damaged zone of seismic fault (Boullier et al., 2004).

**2.2 Damage dynamics.**
The mechanical loading of rocks involves local inelastic processes that produce acoustic wave emissions (AE). This provides a tool for following the damage dynamics during the deformation and failure processes. The correlation between AE activity and macroscopic inelastic strain has been established in many experimental (e.g. Lockner and Byerlee, 1991; Scholz, 1968a) and numerical (e.g. Young et al., 2000) studies. The AE tool has been extensively used at the laboratory rock sample scale (see Lockner, 1993 for a review) and at an intermediate scale between the lab scale and the large tectonic earthquakes, for studies of seismicity and rockburst in mines or tunnels (e.g. Nicholson, 1992; Obert 1977) and for monitoring slope stability related to either open mine, quarry, landslides, rocky-cliffs or volcano flanks (Amitrano et al., 2005; Hardy and Kimble, 1991; e.g. McCauley, 1976).
Fracturing dynamics during mechanical loading, observed through AE monitoring, usually displays a power law distribution of acoustic events size (Scholz, 1968b).

$$N(>A) \sim A^{-b} \qquad (1)$$

Where $A$ is the maximum amplitude of AE events, $N(>A)$ is the number of events with maximum amplitude greater than $A$, $b$ is a constant. In a log-log representation, this distribution appears linear and b is given by the slope of the line.

$$\log N(>A) \sim -b \log A \qquad (2)$$

This distribution exhibits remarkable similarity to the Gutenberg-Richter relationship observed for earthquakes (Gutenberg and Richter, 1954).

$$\log N(>M) \sim -bM \qquad (3)$$

where $N(>M)$ is the number of earthquakes with a magnitude larger than $M$. Assuming that the magnitude is proportional to the log of the maximal amplitude of the seismic signal, the $b$ values obtained from the magnitude or the amplitude can be compared (Weiss, 1997). Rigorously, the amplitude measured at a given distance from the source should be corrected for the attenuation. Nevertheless, theoretical (Weiss, 1997) and experimental studies (Lockner, 1993) have shown that attenuation has no significant effect on the $b$ value.
In order to quantitatively estimate the damage localization, many authors use the spatial correlation integral method (Grassberger and Prococcia, 1983) for characterizing the distribution of the AE source cloud. The spatial correlation integral is defined as:

$$C(r) = \frac{2}{N(N-1)} N(R>r) \qquad (4)$$

where $N$ is the total number of damage events, $N(R > r)$ is the number of pairs of events separated by a distance smaller than r. If this integral exhibits a power law, $C(r) \sim r^{D_2}$, the population can be considered as fractal and $D_2$ is the correlation dimension.

As power laws indicate scale invariance and because of the similarities in the physics of the phenomena (wave propagation induced by fast source motion), AE of rocks observed in the laboratory has been considered as a small scale model for the seismicity in rock masses (rockbursts) or in the Earth's crust (earthquakes) (Mogi, 1962; Scholz, 1968b). Observations of both earthquakes and AE show variations of the $b$ value in time and space domains which are usually explained using fracture mechanics and/or the self organized criticality (SOC) concept. Mogi (1962) suggested that the $b$-value depends on material heterogeneity, a low heterogeneity leading to a low $b$-value. Scholz (1968b) observed that the $b$-value decreases before the peak stress is achieved and argued for a negative correlation between $b$-value and stress. Main et al. (1989) observed the same variation but invoked a negative correlation between the $b$-value and the stress intensity factor $K$. Following this idea, the authors proposed different patterns of $b$-value variation before the macrorupture, driven by fracture mechanics and the type of rupture (brittle-ductile). The relationship between the $b$-value and the fractal dimension $D_2$ of AE source locations was also investigated (Lockner and Byerlee, 1991; Lockner et al., 1991) and showed a decrease of b value simultaneous to the strain localization, i.e. to a decrease of the $D_2$ value which appear to be associated with a ductile macroscopic behavior. Numerical models based on elastic damage (Tang, 1997; Tang and Kaiser, 1998) succeed in simulating brittle behavior. Discrete element models simulating macroscopic behavior ranging from brittle to ductile and power law distributions of earthquakes have also been proposed (Li et al., 2000; Place and Mora, 2000; Wang et al., 2000). Wang et al. (2000) argue that the $b$-value depends on the cracks density distribution but do not report a relation between the $b$-value and the type of mechanical behavior. Amitrano et al. (1999) and Amitrano (2003) proposed a model which simulates both ductile and brittle behavior and show that the $b$-value depends on the macroscopic behavior. These results suggest that a relationship between the $b$-value and the macroscopic behavior may exist.

Mori and Abercombie (1997) observed a decrease of the $b$-value with increasing depth for earthquakes in California. They suggested that the $b$-decrease was related to a diminution of the heterogeneity as depth increases. Systematic tests of the dependence of the $b$ value on depth have been performed by Gertenberger (2001) which confirm these general results but show some discrepancies depending on the tectonic stress regime. The depth dependence of the $b$-value have also been observed for the western Alps seismicity (Sue et al., 2002) and for earthquakes sequence along the Aswan Lake in Egypt (Mekkawi et al., 2002). Recent results show that the $b$-value depends on the tectonics regime (Schorlemmer and Wiemer, 2005) and systematically variate for normal faulting (extension), inverse faulting (compression) and strike-slip (plane shearing). Other authors have used cellular automata (Chen et al., 1991; Olami et al., 1992) or lattice solid models (Zapperi et al., 1997b) to simulate power law distribution of avalanches.

Acoustic monitoring has been used for studying the damage acceleration before failure. Laboratory scale experiments on heterogeneous material have revealed that the acceleration follows a power law (Guarino et al., 1998; Johansen and Sornette, 2000; Nechad et al., 2005).

This power law accelerating microdamage before the macroscopic brittle failure has been suggested to be the fingerprint of a critical behaviour analog to a thermodynamics phase transition (Buchel and Sethna, 1997; Kun and Herrmann, 1999; Sornette, 2000; Sornette and Andersen, 1998; Zapperi et al., 1997a). This kind of acceleration has been recently reported for the seismic events preceding the collapse of a chalk cliff (Amitrano et al., 2005).

Nonetheless many other experiments do not reproduce the patterns predicted by statistical physics before brittle failure and the applicability of these brittle failure models to the earth crust fracturing is still debated, e.g. the so-called critical point hypothesis for earthquakes (Bufe and Varnes, 1993; Jaume and Sykes, 1999; Zoller and Hainzl, 2002).

## 3. Numerical modelling
### 3.1 Progressive damage model

The model we use here (Amitrano, 2003; Amitrano et al., 1999) integrates the main features of two previous models which simulate respectively macroscopic ductility (Zapperi et al., 1997b) or brittleness (Tang, 1997; Tang and Kaiser, 1998). It is based on progressive isotropic elastic damage. The effective elastic modulus, $E_{eff}$, is expressed as a function of the initial modulus, $E_0$ and damage, $D$.

$$E_{eff} = E_0(1-D) \quad (5)$$

Such a relation works when the considered volume is large compared with the defect size, such as cracks, and then can be considered as a mesoscale relationship. The damage parameter, $D$, has been proposed to be related to crack density (see Kemeny and Cook, 1986 for a review). The simulated material is discretized using a 2 D finite element method with plane strain hypothesis. The loading consists in increasing the vertical displacement of the upper model boundary. When the stress in an element exceeds a given damage threshold, its elastic modulus is multiplied by a factor *(1-D)*, *D* being constant. Because of the elastic interaction, the stress redistribution around a damaged element can induce an avalanche of damages that we call an event. The total number of damaged elements during a single loading step is the avalanche size, which is comparable to the acoustic emission event size observed in laboratory experiments.

The Mohr-Coulomb criterion is used as a damage threshold,

$$\tau = \mu\sigma + C, \quad (6)$$

where $\tau$ is the shear stress; $\sigma$ is the normal stress; $C$ is the cohesion; and $\mu$ is the internal friction coefficient. This criterion has been chosen because of its simplicity and because it allows us to check the sensitivity of the model to each parameter ($C$, $\mu$, $\sigma$) in an independent manner. In the absence of heterogeneity, the behavior of the model is entirely homogenous, (i.e. no damage localization occurs) and the local behavior is replicated at the macroscopic scale. To obtain macroscopic behaviors differing from those of the elements and damage localization it is necessary to introduce heterogeneity. To simulate material heterogeneity the cohesion of each element, $C$, is randomly drawn from a uniform distribution.

### 3.2 Numerical simulation results

The study of the model sensitivity has shown that confining pressure, cohesion and damage parameter D do not change the type of macroscopic behaviour, nor the kind of localization mode, but only the macroscopic stress level (Amitrano et al., 1999). On the contrary, the internal friction influences both the macroscopic behaviour and the damage localisation. Consequently, we present the results obtained with different values of µ and fixed values for the others parameters: $E_{initial}$ = 50 GPa, $\nu$ = 0.25, C random between 25 and 50 MPa, D = 0.1. The simulations are performed with uniaxial stress conditions ($\sigma_2=\sigma_3=0$).

Figure 1 presents the macroscopic behaviour simulated for different values of the internal friction $\mu$. $\sigma_1$ is the major main stress and $\varepsilon_1$ is the major main strain calculated at the model boundary, i.e., the mean values over the boundary. One may observed that for $\mu$ ranging from 0 to 1.5, which corresponds to the variation range observed for rocks, the macroscopic behaviour ranges from pure ductility to brittleness, without changing the elementary behaviour of the elements. Figure 2 displays the damage map at the end of the simulations for 3 simulations with $\mu$=0, $\mu$=0.4 and $\mu$=1. The color scale bar indicates the damage of each element (i.e. $E/E_0$). One may observe that damage localization is dramatically enhanced when the $\mu$ parameter increases.

In order to estimate quantitatively the grade of damage localization we calculated the correlation integral of the damage at the end of each simulation. The correlation dimension $D_2$ was calculated by linear regression in the log-log plane. The range used for the linear regression was restricted to the linear part of the curves (r=0.01-0.1). A value of $D_2$ near 2 indicates that the spatial repartition is homogeneous. A value near 1 indicates that the damage is localized along a line. Figure 3 shows the correlation integrals calculated for all the simulations and the corresponding value of the correlation dimension. These results show that the damage localization progressively increases as the µ parameter is increased. The study of the damage localization during the simulation shows, in the case of brittle behaviour (i.e., presence of a stress drop after the stress peak), the band localization occurs during the macrofailure. This is in good agreement with experimental studies using acoustic emission to assess the damage localization process (Lockner and Byerlee, 1991). The ductile behavior is associated with a diffuse damage ($D_2 \sim 2$). Note that tuning the $\mu$ parameter allows us to simulate all intermediary behaviours from pure ductility to pure brittleness. This progressive change in the macroscopic behaviour is associated with a progressive change from diffuse to localized damage.

The brittle-ductile behavior appears to be related to the diffuse-localized damage repartition respectively. These results suggest that the brittle-ductile transition and the associated localized-diffuse transition are controlled by a unique parameter, the internal friction $\mu$. This is in agreement with experimental results for which it has been established that materials with large internal friction tend to fail by localized failure, whereas those with very low internal friction angle fail by a diffuse mode (Jaeger and Cook, 1979). Because we use the Mohr-Coulomb criterion, for which the internal friction angle is independent of the confining pressure, the simulated macroscopic behavior is insensitive to the confining pressure. This feature is in disagreement with experimental observations, which demonstrate that the increase of confining pressure induces the brittle-ductile transition. An improvement of the model has been proposed (Amitrano, 2003) using a pressure sensitive criterion in order to simulate the pressure induced brittle-ductile transition. As we focus here on the failure induced by damage accumulation, all the simulations presented here are realized with the simplest first version of the model.

In order to better quantify the spatial structure of the damage, we calculated the directional spatial correlogram (*DSC*) of the total amount of damage $D=1-E/E_0$. For a given direction, $\vec{n}$, the *DSC* is calculated as the autocorrelation function along this direction, i.e. the correlation between the damage value observed at point *x* and at point *x'* separated by a distance $\lambda$ along direction $\vec{n}$ (at an angle α relatively to the loading direction). The correlation is calculated as the covariance between $D(\vec{x})$ and $D(\vec{x}+\lambda\vec{n})$ divided by the variance of $D(\vec{x})$.

$$DSC(\alpha,\lambda) = \frac{\text{var}(D(\vec{x}), D(\vec{x}+\lambda\vec{n}))}{\text{var}(D(\vec{x}))} \quad (7)$$

We calculated the *DSC* as a function of the distance $\lambda$ for all values of $\alpha$ between 0 and 180°, with a step of 5°. This analysis is able to reveal the spatial correlation of the damage and its anisotropy. The direction of the damage band is characterized by a long range correlation and the perpendicular direction by a correlation length equivalent to the band thickness. Figure 4 shows the *DSC* for three different simulations performed with $\mu=0$, $\mu=0.4$ and $\mu=1.5$ respectively. *DSC* was calculated for 10 successive steps of the simulation corresponding to an equal number of damage events. The legend indicates the corresponding normalized deformation. The direction $\alpha$ corresponds to the direction of the shear band and $\alpha+\pi/2$ to the shear band normal. One may observe that the correlation length at the beginning of the simulation is near zero for directions $\alpha$ and $\alpha+\pi/2$. This is observed for all the directions, indicating an isotropic damage with no spatial correlation. The damage at a given point is independent of the damage in its neighbourhood. As the simulation progress, the correlation length increases in the same manner for all the directions. At a given step the anisotropy appears as the correlation length increases faster in the direction of the future shear band. In the perpendicular direction the correlation becomes negative for a length corresponding to the thickness of the shear band. This progression from isotropic uncorrelated damage to anisotropic correlated damage is observed for all the simulations. An interesting point is that the increase of correlation length appears significantly before the peak stress, including for the brittle behavior. This should be used for the forecasting of macroscopic failure.

The damage event size (i.e. the number of damaged elements in each single avalanche) distribution has been analyzed as a function of the internal friction *μ* and of the deformation. Figure 5 shows the cumulative distribution function (cdf) and the probability density function (*pdf*) of the damage event size for simulations performed with various *μ*. The *cdf* and *pdf* show power law trends in the range 1-100. For larger size events, *cdf* displays a cut-off (lack of large events compared to the power-law trend) for low values of *μ*. As *μ* increases this cut-off progressively vanishes. For larger values of *μ* the cut-off is replaced by an excess of large events compared to the power-law trend. The larger event corresponds in this last case to the macro failure event.

Figure 6 shows the evolution of *cdf* during the deformation for two simulations with *μ*=0 and *μ*=1.5 respectively. For both simulations the *cdf* displays a power law trend, for low size events with a decrease of the exponent. For *μ*=0, this decrease is associated with a cut-off for large sizes. For *μ*=1.5 the decrease is of lesser amplitude and no cut-off appears. On the contrary an excess of large events appears in the period including the macro failure which is out of trend compared with the power law.

**4. Summary and discussion**

In the first part of the paper we have reviewed works related to the observation of rupture in rocks which reveal complex patterns including fractal properties of damage structure, as self affinity of fracture surfaces and power-law distribution of grain size in highly damaged zones. The damage dynamics, as observed by acoustic emission, displays power-law distributions with exponent value depending on the pressure and on the proximity of the failure. The final failure of the material appears to be the result of damage localization / accumulation. For investigating this process we used a simple progressive damage model able to reproduce the major part of these observations.

The proposed model is based on an elementary progressive damage within an elastic heterogeneous model. Each element has isotropic properties associated with a simple behaviour, i.e., decrease of the elastic modulus by discrete damage events. At the macroscopic scale, the model reproduce different aspects of a complex behaviour: the mechanical behaviour is non-linear and ranges from ductile to brittle, the final damage state has a fractal structure, the size-frequency of damage events follows a power-law. As these properties are not incorporated at the elementary scale, they are emerging properties of the system due to the interaction between elements. According to that, the simulated deformation process can be considered as a complex system. The emergence of scale free distributions for both size and space distributions is a supplementary aspect of this complexity. We have shown that changing the internal friction $\mu$ modifies all the macroscopic properties. In particular we have observed that the damage localization is dramatically enhanced when the $\mu$ parameter increases. The study of the stress field around a single defect (Amitrano et al., 1999), has shown that this parameter strongly influences the interaction geometry between elements. The higher the $\mu$ parameter is, the more anisotropic the interaction is. This low scale anisotropic interaction controls the mode of damage localisation we can observe at the macroscopic scale. In the case of a highly localized damage (i.e. for $\mu > 1$), the localization occurs in an instable mode associated with a dramatic stress drop, we considered as the fingerprint of a macroscopically brittle behaviour. The $\mu$ parameter influences also the event size distribution as a result of the local interaction between elements (more or less anisotropic). Hence the small scale anisotropy influences the damage localisation, the avalanches dynamics and the macroscopic behaviour. Despite the limited scale dynamics of the model, we observed the scaling relationship for damage structure, over 1.5 order of magnitude, characterized by a power-law trend of the spatial correlation integral. The event size distribution is a power-law which evolves during the simulation showing a decrease of the exponent, i.e. an increase of the event mean size, in agreement with laboratory observation. In the brittle case, the failure corresponds to the larger event in size, which is out of range compared with the power-law trend. The study of spatial correlation of the damage during the simulation has shown an increase of the correlation length more pronounced in the direction of the localization band. This increase of length is associated with the event size increase. The failure occurs when correlation length becomes large enough for leading to a macrofailure event, i.e. the size of the model.

These numerical results show that damage accumulation leading to the failure is strongly influenced by the local interaction geometry which, depending on the anisotropy, can lead either to a macroscopically ductile behaviour with progressive localization, i.e., without stress drop, or to a macroscopically brittle behaviour associated to a sudden localization event. All these results are obtained without changing the basic elementary behaviour but only the interaction. This mesoscale approach could be an alternative to the microscopic approach, dedicated to the study of fracture propagation and to the macroscopic approach based on constitutive laws. It provides observables which are emerging properties from elementary interaction and, in this regard, consider the rupture process as a complex phenomenon.

**5. Conclusions**
We have shown that the deformation and rupture process in rocks reveals many complex behaviours as fractal structure of damage, power-law distribution of damage events, damage localization associated to brittleness/ductility. Using a simple model based on elastic interaction and progressive damage we succeed in reproducing the major part of this complexity. The study of damage spatial correlation revealed that the process of damage localization is related to the passage from isotropic uncorrelated damage to anisotropic correlated damage. The differences in the localization mode (more or less diffuse and progressive) are related to the type of local interaction between elements which can be tuned, in the model we used, by changing the internal friction. Such a mesoscopic approach could be an alternative to the microscopic one based on fracture mechanics concept or to the macroscopic one based on constitutive laws.

# 6 References

Amitrano, D. (2003). Brittle-ductile transition and associated seismicity: Experimental and numerical studies and relationship with the b-value. *J. Geophys. Res.*, *108*(B1), 2044, doi:10.1029/2001JB000680.

Amitrano, D., Grasso, J.-R. and Hantz, D. (1999). From diffuse to localized damage through elastic interaction. *Geophysical Research Letters*, *26*(14), 2109-2112.

Amitrano, D., Grasso, J.R. and Senfaute, G. (2005). Seismic precursory patterns before a cliff collapse and critical-point phenomena. *Geophysical Research Letters*, *32*(8), L08314, doi:10.1029/2004GL022270.

Amitrano, D. and Schmittbuhl, J. (2002). Fracture roughness and gouge distribution of a granite shear band. *J. Geophys. Res.*, *107*(B12, 2375, doi:10.1029/2002JB001761), 16 p.

Biarez, J. and Hicher, P.-Y. (1997). Influence de la granulométrie et de son evolution par rupture de grains sur le comportement mécanique des materiaux granulaires. *Revue francaise de genie civil*, *1*(4), 607-631.

Bouchaud, E. (1997). Scaling properties of cracks. *J. Phys. Condens. Matter*, *9*, 4319.4344.

Bouchaud, E., Lapasset, G. and Plane`s, J. (1990). Fractal dimension of fractured surfaces: A universal value? *Europhys. Lett.*, *13*, 73-79.

Boullier, A.M., Fujimoto, K., Ito, H., Ohtani, T., Keulen, N., Fabbri, O., Amitrano, D., Dubois, M. and Pezard, P. (2004). Structural evolution of the Nojima fault (Awaji Island, Japan) revisited from the GSJ drillhole at Hirabayashi. *Earth, Planets and Space*, *56*, 1233-1240.

Brown, S.R. and Scholz, C.H. (1985). Broad bandwith study of the topography of natural rock surface. *Journal of Geophysical Research*, *90*(B14), 12575-12582.

Buchel, A. and Sethna, J.P. (1997). Statistical mechanics of cracks: Fluctuations, breakdown, and asymptotics of elastic theory. *Phys. Rev. E*, *55*(6), 7669-7690.

Bufe, C.G. and Varnes, D.J. (1993). Predictive modeling of the seismic cycle and of the greater San francisco Bay region. *J. Geophys. Res.*, *98*, 9871-9883.

Chen, K., Bak, P. and Obukhov, S.P. (1991). Self-organized criticality in a crack-propagation model of earthquakes. *Physical Review A*, *43*(2), 625-630.

Costin, L.S. (1983). A microcrack model for the deformation and failure of brittle rock. *Journal of Geophysical Research*, *88*(B11), 9485-9492.

Cowie, P., Vanneste, C. and Sornette, D. (1993). Statistical physics model for the spatiotemporal evolution of faults. *Journal of Geophysical Research*, *98*(B12), 21809-21821.

Cox, S.J.D. and Meredith, P.G. (1993). Microcrack formation and material softening in rock measured by monitoring acoustic emission. *Int. J. Rock Mech. Min. Sci. and Geomech. Abstr.*, *30*(1), 11-24.

Gerstenberger, M., Wiemer, S. and Giardini, D. (2001). A systematic test of the hypothesis that b varies with depth in California. *Geophysical Research Letters*, *28*(1), 57-60.

Grassberger, P. and Prococcia, I. (1983). Measuring the strangeness of stange attractors. *Physica*, *9*, 189-208.

Guarino, A., Garcimartin, A. and Ciliberto, S. (1998). An Experimental Test of the Critical Behaviour of Fracture Precursors. *European Physical Journal B*, *6*(1), 13-24.

Gutenberg, B. and Richter, C.F., 1954. Seismicity of the earth and associated phenomena. Princeton University Press, Princeton.

Hardy, H.R. and Kimble, E.J. (1991). Application of high-frequency AE/MS techniques to rock slope monitoring. In: Hardy (Editor), Vth Conf. AE/MS Geol. Str. and Mat. Trans Tech Publication, Germany, The pennsylvania State University, 457-477.

Hirata, T., Satoh, T. and Ito, K. (1987). Fractal structure of spacial distribution of microfracturing in rock. *Geophys. J. R. astr. Soc.*, *90*, 369-374.

Jaeger, J.C. and Cook, N.G.W., 1979. Fundamentals of rock mechanics. Chapman and Hall, London, 593 pp.

Jaume, S.C. and Sykes, L.R. (1999). Evolving Towards a Critical Point: A Review of Accelerating Seismic Moment/Energy Release Prior to Large and Great Earthquakes. *Pageoph*, *155*, 279-305.

Johansen, A. and Sornette, D. (2000). Critical ruptures. *Eur. Phys. J. B*, *18*, 163-181.

Keller, J.V.A., Hall, S.H. and McClay, K.R. (1997). Shear fracture pattern and microstructural evolution in transpressional fault zones from field and laboratory studies. *Journal of Structural Geology*, *19*(9), 1173-1187.

Kemeny, J. and Cook, N.G.W. (1986). Effective moduli, Non-linear deformation and strength of a cracked elastic solid. *Int. J. Rock Mech. Min. Sci. and Geomech. Abstr.*, *23*(2), 107-118.

King, G.C.P. and Sammis, C.G. (1992). The mechanims of finite brittle strain. *Pageoph*, *138*(4), 611-640.

Kranz, R.L. (1983). Microcracks in rocks: a review. *Tectonophysics*, *100*, 449-480.

Kun, F. and Herrmann, H.J. (1999). Transition from damage to fragmentation in collision of solids. *Phys. Rev. E*, *59*, 2623-2632.

Li, H.L., Bai, Y.L., Xia, M.F., Ke, M.F. and Yin, X.C. (2000). Damage localisation as a possible precursor of earthquake rupture. *Pageoph*, *157*, 1929-1943.

Lin, A. (1999). S-C cataclasite in granitic rock. *Tectonophysics*, *304*, 257-273.

Lockner, D.A. (1993). The role of acoustic emission in the study of rock fracture. *Int. J. Rock Mech. Min. Sci. and Geomech. Abstr.*, *30*(7), 883-899.

Lockner, D.A. and Byerlee, J.D. (1991). Precursory AE patterns leading to rock fracture. In: Hardy (Editor), Vth Conf. AE/MS Geol. Str. and Mat. Trans Tech Publication, Germany, The pennsylvania State University, 45-58.

Lockner, D.A., Byerlee, J.D., Kuskenko, V., Ponomarev, A. and Sidorin, A. (1991). Quasi-static fault growth and shear fracture energy in granite. *Nature*, *350*, 39-42.

Main, I.G., Meredith, P.G. and Jones, C. (1989). A reinterpretation of the precursory seismic b-value anomaly from fracture mechanics. *Geophysical Journal*, *96*, 131-138.

Mair, K., Main, I. and Elphick, S. (2000). Sequential growth of deformation bands in the laboratory. *Journal of Structural Geology*, *22*, 25-42.

Marone, C. and Scholz, C.H. (1989). Particle-size distributions and microstructures within simulated fault gouge. *Journal of Structural Geology*, *11*(7), 799-814.

McCauley, M.L. (1976). Microsonic detection of landslides, 54th T.R.B annual meeting . Transportation Research Record 581, Washington DC, 25-30.

Mekkawi, M., Hassoup, A., Grasso, J.R. and Schnegg, P.A. (2002). Fractal and spectral analysis of the earthquake sequences along the Aswan Lake, Egypt, 26th General Assembly of the European Geophysical Society, Nice, abstract n°EGS02-A-00027.

Menendez, B., Zhu, W. and Wong, T.-F. (1996). Micromechanics of brittle faulting and cataclastic flow in Berea Sandstone. *Journal of Structural Geology*, *18*(1), 1-16.

Michibayashi, K. (1996). The role of intragranular fracturing on grain size reduction in feldspar during mylonitization. *Journal of Structural Geology*, *18*(1), 17-25.

Mogi, K. (1962). Magnitude frequency relations for elastic shocks accompanying fractures of various materials and some related problems in earthquakes. *Bull. Earthquake Res. Inst. Univ. Tokyo*, *40*, 831-853.

Moore, D.E., Summers, R. and Byerlee, J.D. (1989). Sliding behavior and deformation textures of heated illite gouge. *Journal of Structural Geology*, *11*(3), 329-342.

Mori, J. and Abercombie, R.E. (1997). Depth dependence of earthquake frequency-magnitude distributions in California: Implication for rupture initiation. *Journal of Geophysical Research*, *102*(B7), 15081-15090.

Nechad, H., Helmstetter, A., Guerjouma, R.E. and Sornette, D. (2005). Andrade and Critical Time-to-Failure Laws in Fiber-Matrix Composites: Experiments and Model. *Journal of Mechanics and Physics of Solids*, *53*(5), 1099-1127.

Nicholson, C. (1992). Recent developments in rockburst and mine seismicity research. In: T.a. Wawersik (Editor), Rock Mechanics. Balkema, Rotterdam, 1079-1086.

Obert , L. (1977). The microseismic method: discovery and early history, 1st Conf. AE/MS Geol. Str. and Mat. Trans Tech Publications., 11-12.

Olami, Z., Feder, H.J.S. and Christensen, K. (1992). Self-organised criticality in a continuous, nonconservative cellular automaton modeling earthquake. *Physical Review Letters*, *68*(8), 1244-1247.

Place, D. and Mora, P. (2000). Numerical simulation of Localisation phenomena in a fault zone. *Pageoph*, *157*, 1821-1845.

Reches, Z. and Lockner, D.A. (1994). Nucleation and growth of faults in brittle rocks. *Journal of Geophysical Research*, *99*(B9), 18159-18173.

Rice, J. (1975). On the stability of dilatant hardening for saturated rock masses. *J. Geophys. Res.*, *80*(11), 1531-1536.

Rudnicki, J.W. and Rice, J. (1975). Conditions for the localization of deformation in pressure-sensitive dilatant materials. *J. Mechanics and Physics of Solids*, *23*, 371-394.

Sammis, C.G. and Biegel, R.L. (1989). Fractals, Fault-Gouge, and Friction. *Pageoph*, *131*(1/2), 255-271.

Scavia, C. (1995). A method for the study of crack propagation in rock structures. *Geotechnique*, *45*(3), 447-463.

Schmittbuhl, J., Gentier, S. and Roux, S. (1993). Field measurements of the roughness of fault surfaces. *Geophysical Research Letters*, *20*(8), 639-641.

Schmittbuhl, J., Schmitt, F. and Scholz, C. (1995). Scaling invariance of crack surface. *Journal of Geophysical Research*, *100*(B4), 5953-5973.

Scholz (1968a). Microfracturing and the inelastic deformation of rocks in compression. *J. Geophys. Res.*, *73*, 1417-1432.

Scholz, C.H. (1968b). The frequency-magnitude relation of microfracturing in rock and its relation to earthquakes. *Bulletin of the Seismological Society of America*, *58*(1), 399-415.

Scholz, H.C., 1990. The mechanics of earthquakes and faulting. Cambridge University Press.

Schorlemmer, D. and Wiemer, S. (2005). Microseismicity data forecast rupture area. *Nature*, *434*, 1086, doi: 10.1038/4341086a.

Schulson, E.M., Illiescu, D. and Renshaw, C.E. (1999). On the initiation of shear faults during brittle compressive failure: A new mechanism. *Journal of Geophysical Research*, *104*(B1), 695-705.

Sornette, D., 2000. Critical phenomena in natural sciences, Chaos, fractals, Self organization and Disorder: Concepts and tools, Berlin, New-York, 432 pp.

Sornette, D. and Andersen, J.V. (1998). Scaling with respect to disorder in time-to-failure. *European Physical Journal B*, *1*, 353-357.

Sue, C., Grasso, J.R., Lahaie, F. and Amitrano, D. (2002). Mechanical behavior of western alpine structures inferred from statistical analysis of seismicity. *Geophysical Research Letters*, *29*(8, doi:10.1029/2001GL014050), 65-1 65-4.

Tang, C.A. (1997). Numerical simulation of progressive rock failure and associated seismicity. *Int. J. Rock Mech. Min. Sci. and Geomech. Abstr.*, *34*(2), 249-261.

Tang, C.A. and Kaiser, P.K. (1998). Numerical simulation of cumulative damage and seismic energy release during brittle rock failure - Part I: Fundamentals. *Int. J. Rock Mech. Min. Sci. and Geomech. Abstr.*, *35*(2), 113-121.

Turcotte, 1992. Fractals and chaos in geology and geophysics. Cambridge University Press, 220 pp.

Velde, B., Moore, D., Badri, A. and Ledesert, B. (1993). Fractal analysis of fractures during brittle to ductile changes. *Journal of Geophysical Research*, *98*(B7), 11935-11940.

Wang, W. and Scholz, C.H. (1994). Wear processes during frictionnal sliding of rocks: a theoretical and experimental study. *Journal of Geophysical Research*, *99*(B4), 6789-6799.

Wang, Y.C., Yin, X.C., Ke, M.F., Xia, M.F. and Peng, K.Y. (2000). Numerical simulation of rock failure and earthquake process on mesoscopic scale. *Pageoph*, *157*, 1905-1928.

Weiss, J. (1997). The role of attenuation on acoustic emission amplitude distributions and b-values. *Bulletin of the Seismological Society of America*, *87*(5), 1362-1367.

Weiss, J. and Gay, M. (1998). Fracturing of ice under compression creep as revealed by a multifractal analysis. *Journal of Geophysical Research*, *103*(B10), 24005-24016.

Wibberley, C.A.J., Petit, J.-P. and Rives, T. (2000). Micromechanics of shear rupture and the control of normal stress. *Journal of Structural Geology*, *22*, 411-427.

Young, R.P., Hazzard, J.F. and Pettitt, W.S. (2000). Seismic and micromechanical studies of rock fracture. *Geophysical Research Letters*, *27*(12), 1767-1770.

Zapperi, S., Ray, P., Stanley, H.E. and Vespignani, A. (1997a). First-order transition in the breakdown of disordered media. *Physical Review Letters*, *78*, 1408-1411.

Zapperi, S., Vespignani, A. and Stanley, E. (1997b). Plasticity and avalanche behaviour in microfracturing phenomena. *Nature*, *388*(14 august 1997), 658-660.

Zoller, G. and Hainzl, S. (2002). A systematic spatiotemporal test of the critical point hypothesis for large earthquakes. *Geophys. Res. Lett.*, *29*(11), 1558, 10.1029/2002GL014856.

**Figures**

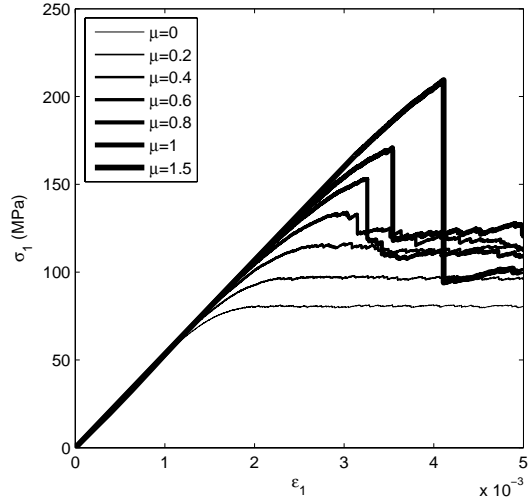

Figure 1 : Macroscopic mechanical behaviours simulated for different $\mu$ values. Tuning the $\mu$ parameter allows switching continuously for ductile to brittle behaviours.

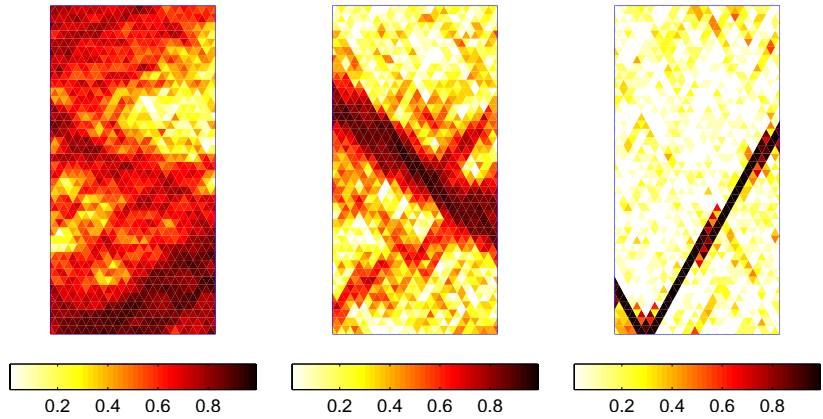

Figure 2 : Damage map for simulation performed with $\mu=0$ (left), $\mu=0.4$ (center), $\mu=1$ (right). The color bar indicates the value of the damage, $D= 1-E/E_0$. The increase of the $\mu$ parameter leads to more localized damage.

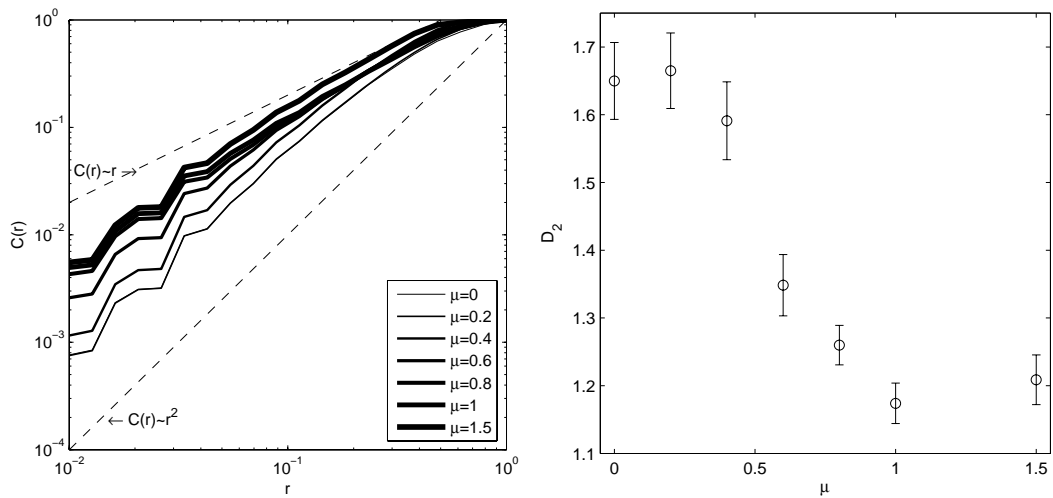

Figure 3: a) Spatial correlation integral, $C(r)$, of the damage events for various values of $\mu$. b) Correlation dimension, $D_2$, calculated by least square regression of $C(r)$ in a loglog plot. The regression is restricted to linear part of $C(r)$, i.e. for $r$=0.01-0.5. Dotted lines indicated particular values of $D_2= 1$ and $D_2=2$.

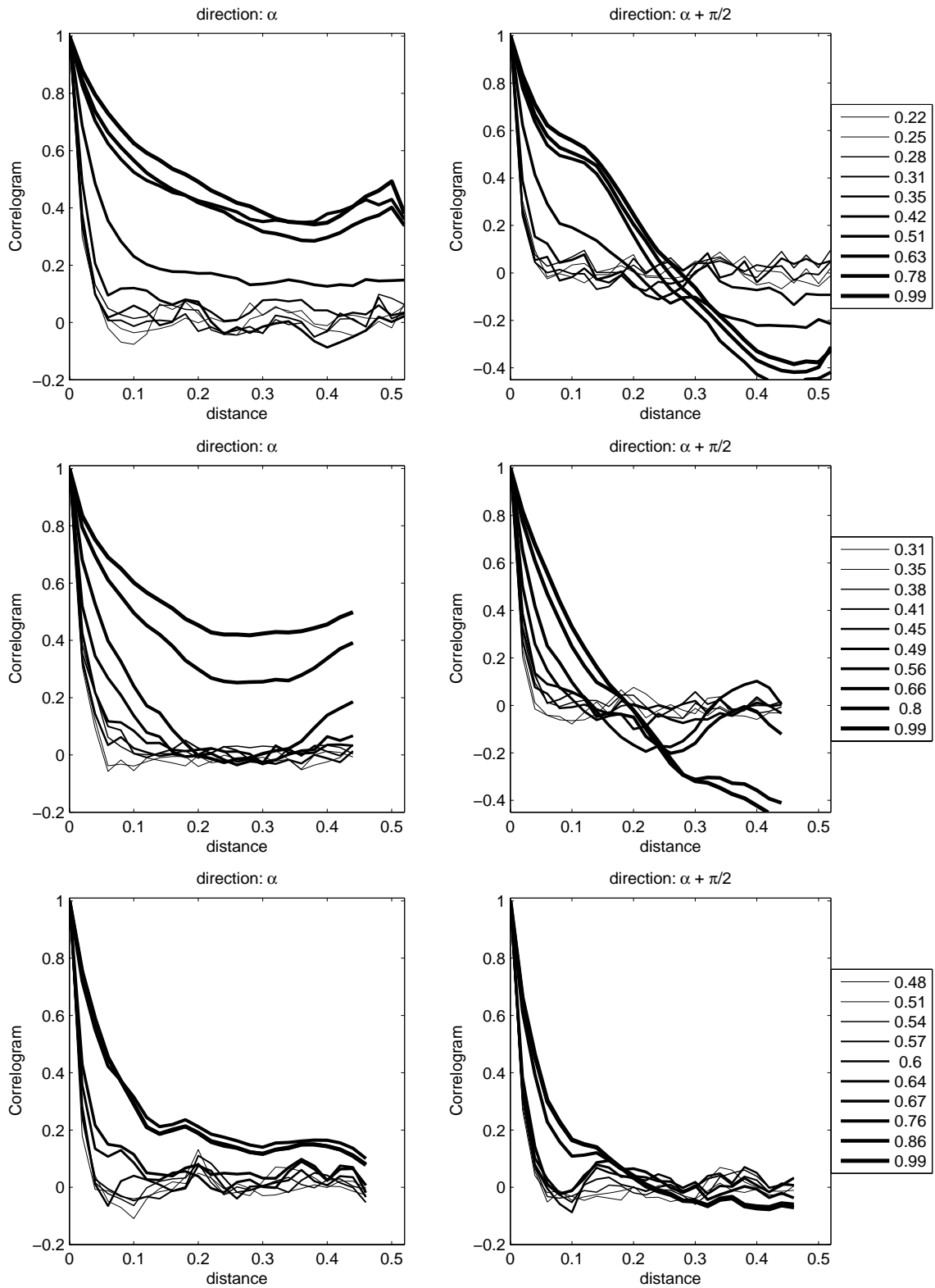

Figure 4: Damage Spatial Correlogram (*DSC*) calculated for three different simulations ($\mu$=0, $\mu$=0.4, $\mu$=1.5). *DSC* is calculated for successive steps. *DSC* is calculated for 10 successive steps of the simulation corresponding to an equal number of damage events. The legend indicates the corresponding normalized deformation. The direction $\alpha$ corresponds to the direction of the shear band and $\alpha+\pi/2$ to the perpendicular.

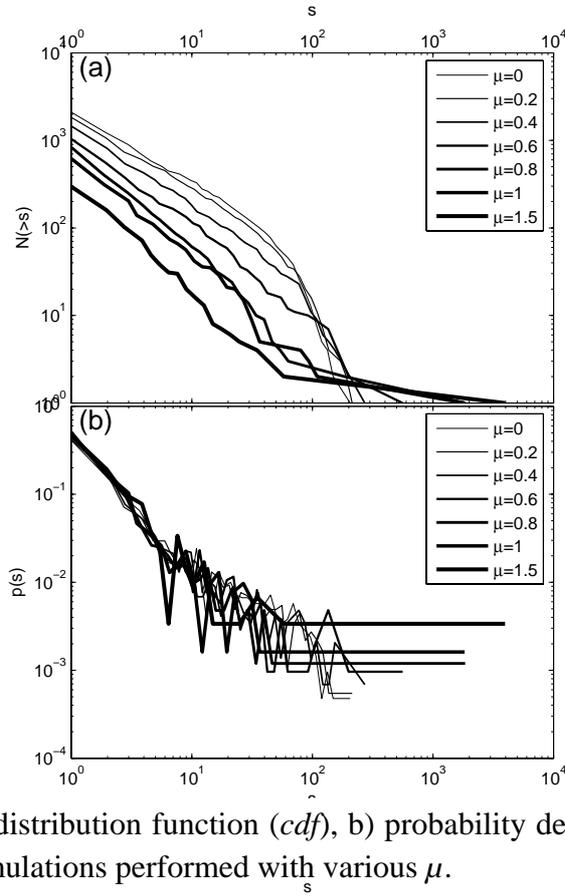

Figure 5: a) Cumulative distribution function (*cdf*), b) probability density function (*pdf*) of the damage event size for simulations performed with various $\mu$.

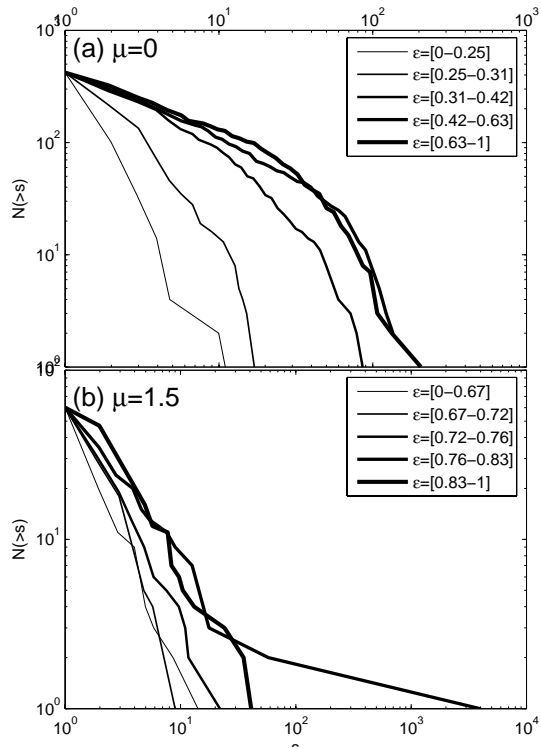

Figure 6: Cumulative distribution function for $\mu=0$ and $\mu=1.5$. The *cdf* has been calculated for 5 successive steps of equal number of events. The legend indicates the corresponding range of normalized deformation.

16